\documentclass{article}
\usepackage{spconf,amsmath,graphicx}
\usepackage{url}
\usepackage{bm}
\usepackage{hyperref}

\title{Embedding a differentiable mel-cepstral synthesis filter \\ to a neural speech synthesis system}
\name{
  \it Takenori Yoshimura$^{1,2}$, Shinji Takaki$^{1,2}$, Kazuhiro Nakamura$^{2}$, Keiichiro Oura$^{2}$,\\
  \it Yukiya Hono$^{1}$, Kei Hashimoto$^{1}$, Yoshihiko Nankaku$^{1}$, and Keiichi Tokuda$^{1,2}$
  \thanks{This work was partly supported by JSPS KAKENHI Grant Number JP22H03614, CASIO Science Promotion Foundation, and Foundation of Public Interest of TATEMATSU.}
}
\address{\\
  $^{1}$Nagoya Institute of Technology, Department of Computer Science, Nagoya, Japan\\
  $^{2}$Techno-Speech, Inc., Department of Research and Development, Nagoya, Japan
}
\begin{document}
\ninept
\maketitle
\begin{abstract}
This paper integrates a classic mel-cepstral synthesis filter into a modern neural speech synthesis system towards end-to-end controllable speech synthesis.
Since the mel-cepstral synthesis filter is explicitly embedded in neural waveform models in the proposed system,
both voice characteristics and the pitch of synthesized speech are highly controlled
via a frequency warping parameter and fundamental frequency, respectively.
We implement the mel-cepstral synthesis filter as a differentiable and GPU-friendly module to enable
the acoustic and waveform models in the proposed system to be simultaneously optimized in an end-to-end manner.
Experiments show that the proposed system improves speech quality from a baseline system maintaining controllability.
The core PyTorch modules used in the experiments will be publicly available on GitHub\footnote{\url{https://github.com/sp-nitech/diffsptk}}.
\end{abstract}
\section{Introduction}
\label{sec:intro}
In statistical parametric speech synthesis~\cite{SPSS}, linear time-variant synthesis filters
such as line spectral pairs~(LSP) synthesis filter~\cite{LSP}, mel-cepstral synthesis filter~\cite{MLSA},
and WORLD vocoder~\cite{WORLD}
are traditionally used as a waveform model to bridge the gap between acoustic features and the audio waveform.
An important advantage of linear synthesis filters is that they can easily control voice characteristics and the pitch of synthesized speech by modifying input acoustic features.
However, the speech quality is often limited due to the linearity of the filters.
In 2016, the first successful neural waveform modeling,
WaveNet~\cite{WaveNet}, was proposed, enabling various nonlinear synthesis filters~\cite{LPCNet,PWG,HiFi,WaveGrad} to have been intensively developed by researchers.
Nonlinear neural synthesis filters outperform linear synthesis filters and can synthesize high-quality speech comparable to human speech.
However, their generation of the waveform which is quite different from training data does not work well,
resulting in less controllability of speech synthesis systems.
The main reason is that the nonlinear synthesis filters ignore the clear relationship between acoustic features and the audio waveform,
which is simply described by conventional linear synthesis filters.
There have been a number of attempts to solve the problem, especially in terms of pitch~\cite{NSF,QPNet,PeriodNet}.
Although they introduced special network structures against neural black-box modeling, the problem has yet to be fully solved.

In this paper, to take the advantage of both linear and nonlinear synthesis filters,
we embed a mel-cepstral synthesis filter in a neural speech synthesis system.
The main advantages of the proposed system are as follows:
1)~Voice characteristics and the pitch of synthesized speech can be easily modified in the same manner as using the conventional mel-cepstral synthesis filter.
2)~Since the embedded filter is essentially differentiable,
speech waveform can be accurately modeled by the simultaneous optimization of the neural waveform model and the neural acoustic model that predicts acoustic features from text.
3)~The neural waveform model is enough to have less trainable parameters because the mel-cepstral synthesis filter captures the rough relationship between acoustic features and the audio waveform.
The proposed system may not require a complex training strategy such as using generative adversarial networks~(GANs)~\cite{PWG} because the rough relationship is already constructed by the embedded filter without training.

The main concern about the proposed system is how to implement the mel-cepstral synthesis filter as a differentiable parallel processing module for efficient model training.
Usually, the mel-cepstral synthesis filter is widely implemented as an infinite impulse response~(IIR) filter~\cite{MLSA}.
This recursive property is not suitable for parallel processing. 
We solve the problem by formulating the mel-cepstral synthesis filter as cascaded finite impulse response~(FIR) filters, \textit{i.e.}, stacked time-variant convolutional layers.

\section{Related work}
Tokuda and Zen~\cite{Direct1,Direct2} combined a neural acoustic model and a cepstral synthesis filter~\cite{LMA}.
The acoustic model was trained by directly maximizing the likelihood of waveform rather than minimizing cepstral distance.
The speech quality was still limited due to the linearity of the cepstral synthesis filter.
LPCNet~\cite{LPCNet} combined an all-pole filter with recurrent neural networks for fast waveform generation.
ExcitNet~\cite{ExcitNet} also used an all-pole filter as the backend of waveform generation.
They did not introduce a specific signal-processing structure into their excitation generation modules.

In 2020, differentiable digital signal processing~(DDSP) has been proposed~\cite{DDSP}.
The main idea is similar to our work: a differentiable linear filter is embedded in a neural waveform model.
One of the main differences is that DDSP assumes a sinusoidal model as a linear synthesis filter.
In the DDSP framework, an acoustic model for speech synthesis is not unified currently~\cite{DDSP2}.
Nercessian~\cite{diffworld} proposed a differentiable WORLD synthesizer for audio style transfer.
The synthesizer uses log mel-spectrogram as a compact representation of spectrum and aperiodicity ratio while our work uses a mel-cepstral representation.

\begin{figure*}[t]
\begin{center}
\vspace{0mm}
\includegraphics[width=1.9\columnwidth]{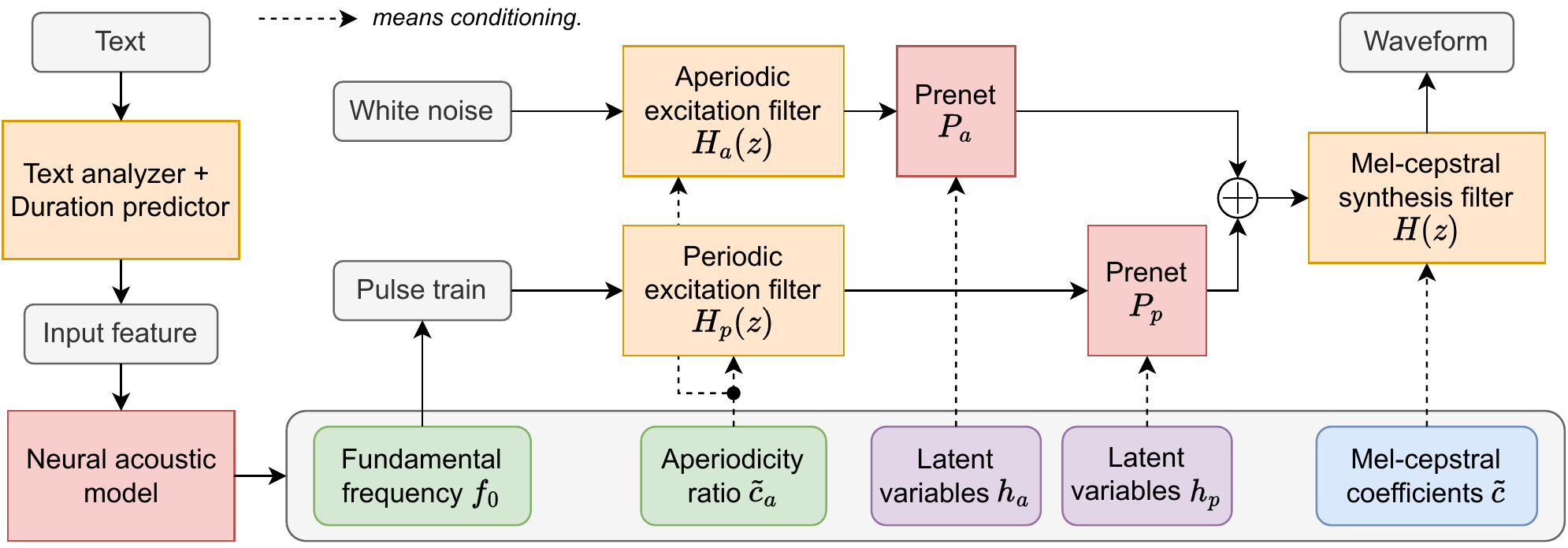}
\vspace{-5mm}
\end{center}
\caption{Neural speech synthesis system embedded mel-cepstral synthesis filter.
Box in red denotes a trainable module and box in orange denotes a non-trainable module.
Text analyzer and duration predictor are preconstructed.
}
\label{fig:overview}
\end{figure*}

\section{Linear synthesis system}
Linear synthesis systems assume the following relation between
an excitation signal $\bm{e} = [e[0], \ldots, e[T-1]]$ and a speech signal $\bm{x} = [x[0], \ldots, x[T-1]]$:
\begin{equation}
X(z) = H(z) E(z),
\label{eq:xhe}
\end{equation}
where $X(z)$ and $E(z)$ are $z$-transforms of $\bm{x}$ and $\bm{e}$, respectively.
The time-variant linear synthesis filter $H(z)$ is parameterized with a compact spectral representation for speech application, \textit{e.g.}, text-to-speech synthesis and speech coding.
In the mel-cepstral analysis~\cite{MCEP}, 
the spectral envelope $H(z)$ is modeled using $M$-th order mel-cepstral coefficients $\{ \tilde{c}(m) \}_{m=0}^M$:
\begin{equation}
H(z) = \exp \sum_{m=0}^M \tilde{c}(m) \tilde{z}^{-m},
\label{eq:hz1}
\end{equation}
where
\begin{equation}
\tilde{z}^{-1} = \frac{z^{-1} - \alpha}{1 - \alpha z^{-1}} 
\end{equation}
is a first-order all-pass function.
Since the scalar parameter $\alpha$ controls the intensity of frequency warping,
the voice characteristics of $\bm{x}$ can be easily modified by changing the value of $\alpha$.
The paper uses Eq.~(\ref{eq:hz1}) as a linear synthesis filter to have the benefits from the property.

Usually, a mixed excitation signal is assumed as the excitation $E(z)$. 
Namely,
\begin{equation}
X(z) = H(z) \{ H_{a}(z) E_{\mathrm{noise}}(z) + H_{p}(z) E_{\mathrm{pulse}}(z) \},
\label{eq:xhee}
\end{equation}
where $E_{\mathrm{noise}}(z)$ and $E_{\mathrm{pulse}}(z)$ are $z$-transforms of white Gaussian noise and pulse train computed from fundamental frequency $f_0$, respectively.
The $H_{a}(z)$ and $H_{p}(z)$ are zero-phase linear filters represented as
\begin{eqnarray}
H_{a}(z) \hspace{-2mm}&=&\hspace{-2mm}
\cos \,\, \sum_{m=0}^{M_a} \,\, \tilde{c}_{a}(m) \tilde{z}^{-m} \\
\hspace{-2mm}&=&\hspace{-2mm}
\exp \!\!\! \sum_{m=-M_a}^{M_a} \!\!\! \tilde{c}'_{a}(m) \tilde{z}^{-m}, \\
H_{p}(z) \hspace{-2mm}&=&\hspace{-2mm}
1 - H_{a}(z),
\end{eqnarray}
where
\begin{eqnarray}
\tilde{c}'_{a}(m) = \left\{ \begin{array}{ll}
\tilde{c}_{a}(0),      & (m = 0) \\
\tilde{c}_{a}(|m|) / 2,  & (m \ne 0)
\end{array} \right.
\end{eqnarray}
and $\{ \tilde{c}_a(m) \}_{m=0}^{M_a}$ is a mel-cepstral representation of aperiodicity ratio~\cite{WORLD}.
The aperiodicity ratio represents the intensity of the aperiodic components of $\bm{x}$ for each frequency bin and its values are in $[0, 1]$.
Equation~(\ref{eq:xhee}) describes a simple relationship between $\bm{x}$ and $\bm{e}$,
but there is a limit to accurate modeling of the speech signal $\bm{x}$.

\section{Proposed method}
To generate speech waveform with high controllability, our approach models speech waveform by extending the conventional linear synthesis filter as
\begin{eqnarray}
X(z) \hspace{-2mm}&=&\hspace{-2mm}H(z) \{
P_a(H_{a}(z) E_{\mathrm{noise}}(z)) \nonumber \\
&&\hspace{8mm} + \, P_p(H_{p}(z) E_{\mathrm{pulse}}(z)) \},
\label{eq:xhp}
\end{eqnarray}
where $P_a(z)$ and $P_p(z)$ are nonlinear filters represented by trainable neural networks called prenets.
To capture speech components that cannot be well captured by acoustic features including mel-cepstral coefficients and $f_0$,
the prenets are conditioned on the two $Q$-dimensional latent variable vectors, $\bm{h}_a$ and $\bm{h}_p$.
The latent variable vectors $\bm{h}_a$ and $\bm{h}_p$ as well as acoustic features are jointly estimated by the acoustic model from text.

\begin{figure*}[t]
\begin{center}
\vspace{0mm}
\includegraphics[width=1.9\columnwidth]{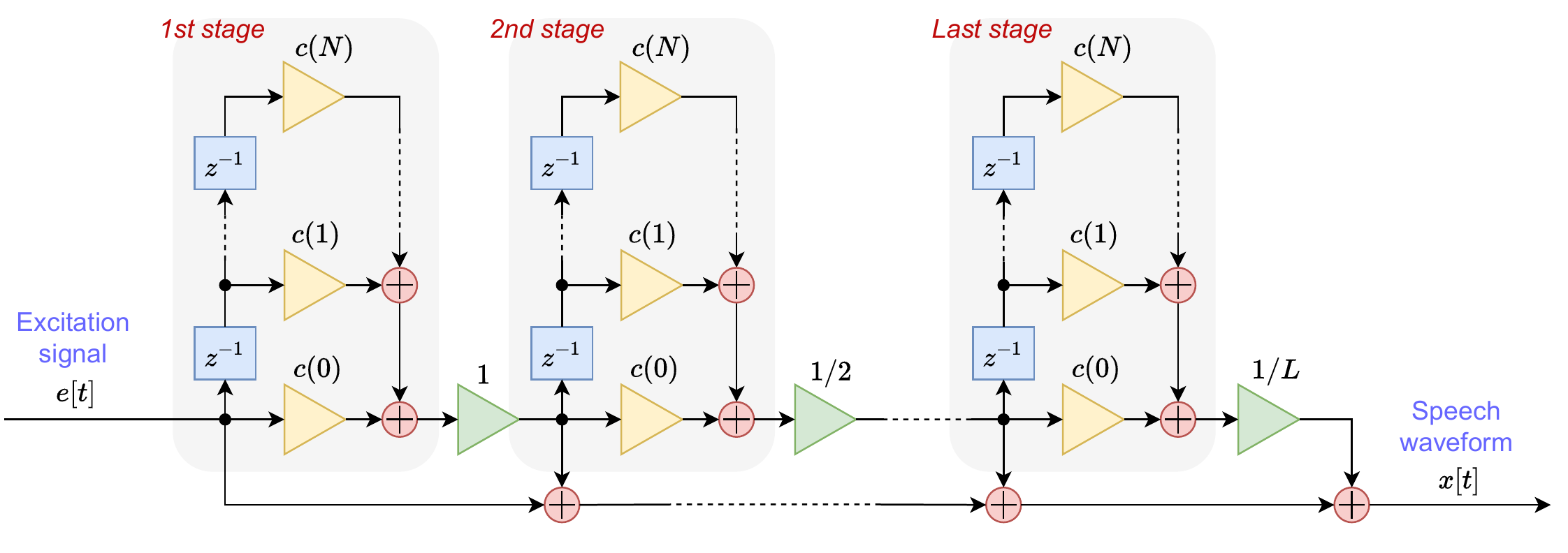}
\vspace{-9mm}
\end{center}
\caption{Mel-cepstral synthesis filter as cascaded FIR filters.}
\label{fig:expdf}
\end{figure*}

An overview of the proposed speech synthesis system is shown in Fig.~\ref{fig:overview}.
Since all linear filters are inherently differentiable,
the acoustic model and prenets can be simultaneously optimized by minimizing the objective
\begin{equation}
\mathcal{L} = \mathcal{L}_{\mathrm{feat}} + \lambda \, \mathcal{L}_{\mathrm{wav}},
\label{eq:L}
\end{equation}
where $\mathcal{L}_{\mathrm{feat}}$ is the loss between the predicted and ground-truth acoustic features,
$\mathcal{L}_{\mathrm{wav}}$ is the loss between predicted and ground-truth speech samples,
and $\lambda$ is a hyperparameter to balance the two losses.
In this paper, the multi-resolution short-time Fourier transform~(STFT) loss~\cite{PWG} is used as $\mathcal{L}_{\mathrm{wav}}$:
\begin{equation}
\mathcal{L}_{\mathrm{wav}} = \frac{1}{2S} \sum_{s=1}^{S} \left(
\mathcal{L}_{\mathrm{sc}}^{(s)}(\bm{x}, \hat{\bm{x}}) + 
\mathcal{L}_{\mathrm{mag}}^{(s)}(\bm{x}, \hat{\bm{x}})
\right),
\end{equation}
where
\begin{eqnarray}
\mathcal{L}_{\mathrm{sc}}^{(s)}(\bm{x}, \hat{\bm{x}}) \hspace{-2mm}&=&\hspace{-2mm}
\frac{\| \, A^{(s)}(\bm{x}) -A^{(s)}(\hat{\bm{x}}) \, \|_F}
     {\| \, A^{(s)}(\bm{x}) \, \|_F}, \\
\mathcal{L}_{\mathrm{mag}}^{(s)}(\bm{x}, \hat{\bm{x}}) \hspace{-2mm}&=&\hspace{-2mm}
\frac{\| \, \log A^{(s)}(\bm{x}) - \log A^{(s)}(\hat{\bm{x}}) \, \|_1}{Z^{(s)}},
\end{eqnarray}
and $\hat{\bm{x}}$ is the predicted speech waveform, $A(\cdot)^{(s)}$ is an amplitude spectrum obtained by STFT under $s$-th analysis condition,
$Z^{(s)}$ is a normalization term,
$\| \cdot \|_F$ is the Frobenius norm, and $\| \cdot \|_1$ is the $L_1$ norm.
Note that during training, ground-truth fundamental frequencies are used to generate the pulse train fed to $H_p(z)$.

\subsection{Differentiable mel-cepstral synthesis filter}
Directly implementing Eq.~(\ref{eq:hz1}) as a digital filter is difficult because it includes the exponential function.
To solve the problem, the mel-log spectrum approximation~(MLSA) filter~\cite{MLSA} has been proposed and is widely used in various applications~\cite{SPSS,LMA}.
The MLSA filter replaces the exponential function with a lower-order rational function using the modified P\'{a}de approximation.
However, the MLSA filter is based on a recursive computation; 
previously computed values as well as the input signal $e[t]$ are used for computing the current value $x[t]$.
This makes training the proposed system with GPU machines very inefficient because they are suitable for parallel computing rather than deep recursive computations.

To solve this, we first convert the mel-cepstral coefficients to cepstral coefficients using a linear transformation~\cite{Freqt}:
\begin{equation}
H(z) \simeq \exp \sum_{m=0}^N c(m) z^{-m},
\label{eq:hz2}
\end{equation}
where $\{ c(m) \}_{m=0}^N$ is the $N$-th order cepstral coefficients computed from $\{ \tilde{c}(m) \}_{m=0}^M$.
Then, by applying the Maclaurin expansion of the exponential function to Eq.~(\ref{eq:hz2}), we obtain
\begin{equation}
H(z) \simeq \sum_{\ell=0}^L \frac{1}{\ell!} \left( \sum_{m=0}^N c(m) z^{-m} \right) ^{\ell} \!.
\label{eq:taylor}
\end{equation}
The infinite series is truncated at the $L$-th term.
Equation~(\ref{eq:taylor}) shows that the mel-cepstral synthesis filter can be implemented as $L$-stage FIR filters,
\textit{i.e.}, $L$ time-variant convolutional layers whose weights are dynamically computed from estimated mel-cepstral coefficients.

Another approach to implement the mel-cepstral synthesis filter is to design a single FIR filter whose filter coefficients correspond to an impulse response converted from the cepstral coefficients $\{ c(m) \}_{m=0}^N$~\cite{c2mpir}.
Note that this approach requires a large GPU memory to handle a filter with many thousands of taps.
Due to GPU machine resources, the paper did not use the latter approach in experiments though it is computationally faster than the former approach described in Eq.~(\ref{eq:taylor}).

\section{Experiment}

\subsection{Experimental setup}
The task to evaluate the proposed system was singing speech synthesis in which pitch controllability is quite important.
We used an internal Japanese corpus containing 70 songs by a female singer: 60 songs were used for training and the remaining were used for testing.
The speech signals were sampled at a rate of 48~kHz.
The acoustic features consisted of 49-th order mel-cepstral coefficients of spectral envelope, log fundamental frequency, 24-th order mel-cepstral coefficients of aperiodicity ratio, and vibrato parameters.
The mel-cepstral coefficients were extracted from the smoothed spectrum analyzed by WORLD~\cite{WORLD} using the mel-cepstral analysis~\cite{MCEP} with $\alpha=0.55$.
The aperiodicity ratio was computed by the TANDEM-STRAIGHT algorithm~\cite{TANDEM} and was also compressed by the mel-cepstral analysis with the same $\alpha$.
The acoustic features were normalized to have zero mean and unit variance.
For the duration predictor in Fig.~\ref{fig:overview},
five-state, left-to-right, no-skip hidden semi-Markov models~\cite{HSMM} were built to obtain the time alignment between the acoustic and score features extracted from the input score.
The input feature for the acoustic model was an 844-dimensional feature vector including the score and duration features.
The acoustic model was the stack of three fully connected layers, three convolutional layers, two bidirectional LSTM layers, and a linear projection layer~\cite{Sinsy}.
For the prenets $P_a$ and $P_p$, the same network architecture of the Parallel WaveGAN generator~\cite{PWG} was used except that the number of channels was reduced by half.
The parameters were optimized through the Adam optimizer~\cite{Adam} with a learning rate of $0.0001$ and mini-batch size of 8.
The models were trained for 260K steps using a single GPU machine with random initialization.
The learning rate was reduced by half to 130K steps.

The dimension $Q$ of $\bm{h}_a$ and $\bm{h}_p$ was 30.
The order of cepstral coefficients $N$ in Eq.~(\ref{eq:hz2}) was 199 for $H(z)$ and 24 for $H_a(z)$.
The order of the Maclaurin expansion $L$ in Eq.~(\ref{eq:taylor}) was set to 20.
For the multi-resolution STFT loss, $S=3$ and the window sizes were set to $\{600, 1200, 2400\}$ where the hop sizes were $80$\% overlap.
Since it is computationally expensive to compute $\mathcal{L}_{\mathrm{wav}}$ for the entire audio waveform, 
we randomly cropped the predicted acoustic feature sequence to 70 frame segments and fed them to the linear filters in the same manner as VITS~\cite{VITS} during training.
The loss in the acoustic feature domain $\mathcal{L}_{\mathrm{feat}}$ in Eq.~(\ref{eq:L}) was
\begin{equation}
\mathcal{L}_{\mathrm{feat}} = 
\frac{1}{1+W} \sum_{w=0}^{W} \mathcal{L}_{\mathrm{nll}}^{(w)}
\left( \bm{o}^{(w)}, \hat{\bm{o}}^{(w)} \right),
\label{eq:feat}
\end{equation}
where
$\bm{o}^{(0)}$ and $\hat{\bm{o}}^{(0)}$ are ground-truth and predicted acoustic features, respectively, 
$(\cdot)^{(w)} (1 \le w)$ denotes their $w$-th order dynamic components, and $W$ was set to 2~\cite{nkazu}.
In Eq.~(\ref{eq:feat}),
\begin{equation}
\mathcal{L}_{\mathrm{nll}}^{(w)} \left( \bm{o}^{(w)}, \hat{\bm{o}}^{(w)} \right) = -\frac{1}{DK} \sum_{k=0}^{K-1}
\mathcal{N} \left(\bm{o}^{(w)}_k \, \Big| \, \hat{\bm{o}}^{(w)}_k, \bm{\varSigma}^{(w)} \right),
\end{equation}
where $D$ is the dimension of the acoustic features,
$K$ is the length of the acoustic features,
and $\bm{\varSigma}^{(w)}$ is a trainable $D$-dimensional time-invariant diagonal covariance matrix.
The parameter $\bm{\varSigma}^{(w)}$ was not used for inference.
The hyperparameter $\lambda$ in Eq.~(\ref{eq:L}) was chosen to be $0.05$.

The methods to compare were as follows:
\begin{itemize}
\item{\textbf{MS-sg}}\hspace{1mm}
A baseline system in which \textbf{sg} means stop gradient.
The acoustic model was trained using only the objective $\mathcal{L}_{\mathrm{feat}}$.
The prenets $P_a$ and $P_p$ were not used.
The waveform was synthesized by feeding the predicted acoustic features to the mel-cepstral synthesis filter described in Eq.~(\ref{eq:taylor}).

\item{\textbf{MS}}\hspace{1mm}
Same as \textbf{MS-sg} except that the acoustic model was trained using the two objectives $\mathcal{L}_{\mathrm{feat}}$ and $\mathcal{L}_{\mathrm{wav}}$.

\item{\textbf{PMS-sg}}\hspace{1mm}
The acoustic model and prenets were simultaneously trained using Eq.~(\ref{eq:L}). 
However, the loss $\mathcal{L}_{\mathrm{wav}}$ propagated through only $\bm{h}_a$ and $\bm{h}_p$ to the acoustic model,
\textit{i.e.}, mel-cepstral coefficients were not explicitly affected by $\mathcal{L}_{\mathrm{wav}}$.

\item{\textbf{PMS}}\hspace{1mm}
The acoustic model and the prenets were simultaneously trained using Eq.~(\ref{eq:L}).

\item{\textbf{PN-sg}}\hspace{1mm}
The acoustic model was trained using only the objective $\mathcal{L}_{\mathrm{feat}}$.
The waveform was synthesized by feeding the predicted acoustic features to the PeriodNet~\cite{PeriodNet}, which is a neural vocoder suitable for singing speech synthesis.
The PeriodNet model was separately trained on the same training data.
Note that the joint training of the acoustic and PeriodNet models is fairly difficult due to computational resources.
\end{itemize}

\subsection{Experimental result}
Two subjective listening tests were conducted.
Both tests evaluated the naturalness of synthesized speech
by the mean opinion score~(MOS) test method., but the second test shifted the pitch up 12 semitones to assess the robustness against fundamental frequencies.
Each of the 12 participants rated 10 phrases randomly selected from the 10 test songs.

\begin{figure}[t]
\begin{center}
\vspace{-2mm}
\includegraphics[width=0.95\columnwidth]{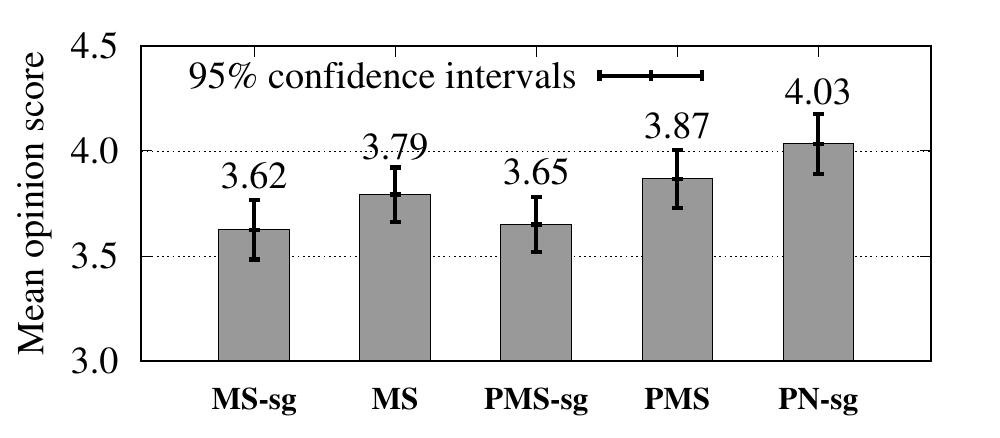}
\vspace{-8mm}
\end{center}
\caption{MOS of naturalness.}
\label{fig:mos1}
\begin{center}
\vspace{-3mm}
\includegraphics[width=0.95\columnwidth]{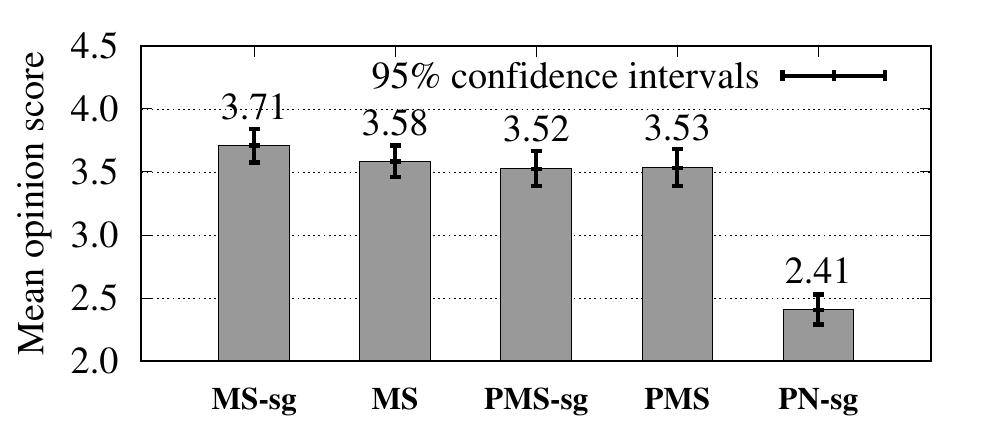}
\vspace{-8mm}
\end{center}
\caption{MOS of naturalness in which input fundamental frequencies were shifted up 12 semitones.}
\label{fig:mos2}
\end{figure}

\begin{figure}[t]
\begin{center}
\vspace{1mm}
\includegraphics[width=1.0\columnwidth]{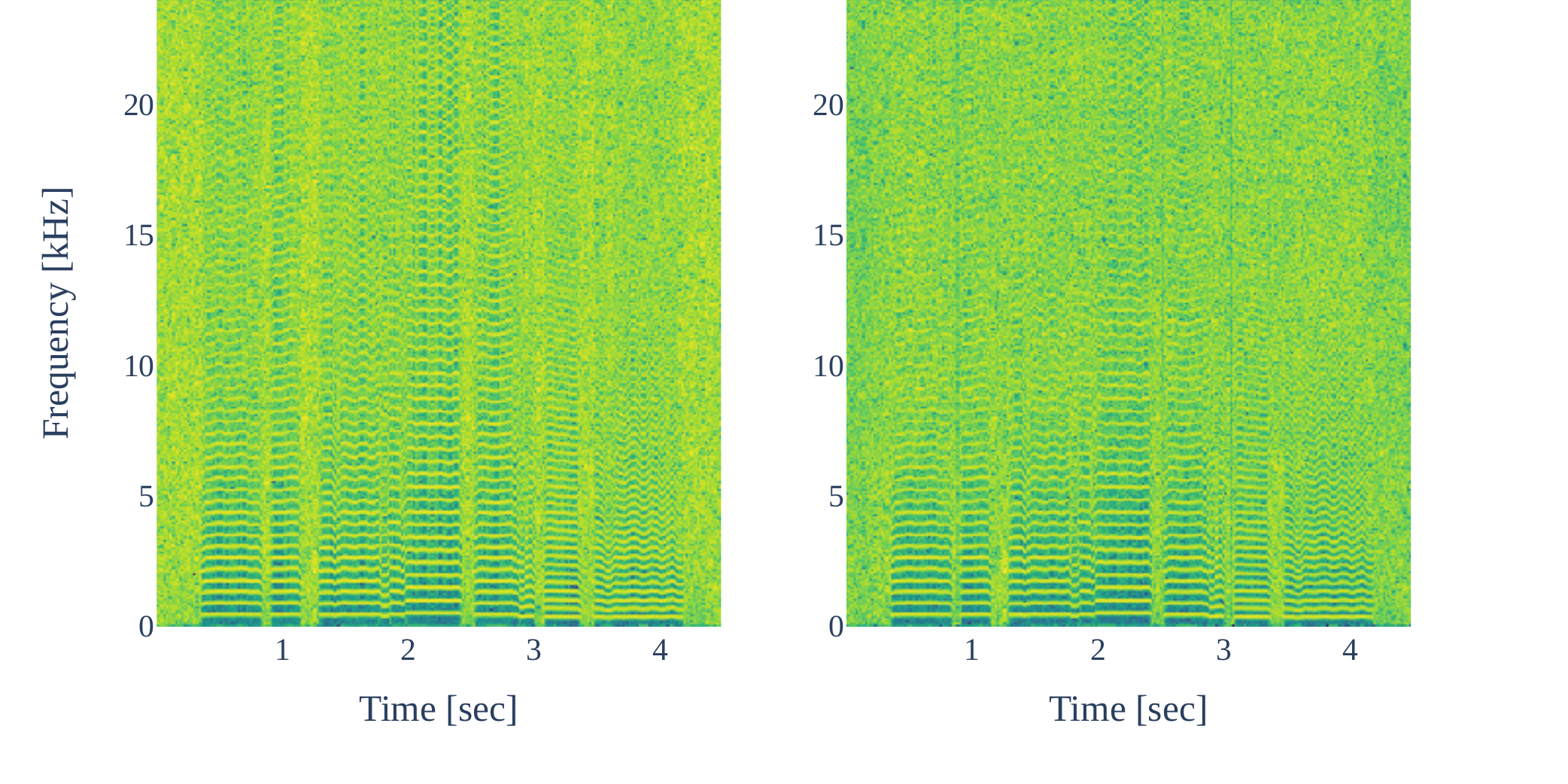}
\vspace{-13mm}
\end{center}
\caption{Example of spectrogram of excitation signal in \textbf{PMS}. Excitation signals are produced w/o prenets (left) and w/ prenets (right).}
\label{fig:spec}
\end{figure}

Figure~\ref{fig:mos1} shows the result of the first test.
\textbf{MS} obtained a higher score than \textbf{MS-sg}.
This means that considering the loss in the waveform domain is effective to improve speech quality.
\textbf{PMS} further improved from \textbf{MS}.
This indicates that the prenets can compensate for the modeling ability of the simple linear filter.
Although \textbf{PMS} did not use any GAN-based training and the network size was smaller than the PeriodNet model, the score was close to \textbf{PN-sg}.
\textbf{PMS-sg} obtained a much lower score than \textbf{PMS}, indicating the importance of the simultaneous optimization.
Figure~\ref{fig:mos2} shows the result of the second test.
Even though \textbf{PMS} includes the prenets, \textbf{PMS} achieved a similar score to that of \textbf{MS}.
This robustness cannot be seen in \textbf{PN-sg}, a pure neural waveform model.
We observed that the proposed system could also control voice characteristics via the frequency warping parameter $\alpha$ 
without degradation.
Audio samples are available on our demo page\footnote{\url{https://www.sp.nitech.ac.jp/~takenori/sample_icassp2023.html}}.
To see the effect of the prenets,
the spectrograms of excitation signals with and without the prenets are plotted in Fig.~\ref{fig:spec}.
The prenets mainly work on the higher frequency domain, which cannot be well captured by mel-cepstral coefficients.

\section{Conclusion}
This paper proposed a neural speech synthesis system with an embedded mel-cepstral synthesis filter.
Experimental results showed that the proposed system can synthesize speech with reasonable quality and has very high robustness against fundamental frequencies.
Future work includes to investigate the optimal network architecture of the prenets and to introduce GAN training to boost the performance of the proposed system.

\vfill\pagebreak

\bibliographystyle{IEEEbib}
\bibliography{strings}

\end{document}